\documentclass[twocolumn]{svjour3}
\usepackage{epsfig}
\usepackage{amsmath}
\usepackage{amssymb}
\usepackage{amsfonts}
\usepackage{mathptmx}
\usepackage{graphicx}
\usepackage{epstopdf}
\usepackage{dcolumn}
\usepackage{multirow}
\usepackage{eucal}
\usepackage{bm}

\begin{document}

\title{Controlling exchange coupling strength in Ni$_x$Cu$_{100-x}$ thin films}

\author{B. Nagy \and Yu. N. Khaydukov \and L. F. Kiss \and Sz. Sajti \and D. G. Merkel \and F. Tanczik\'{o}
\and A. S. Vasenko \and R. O. Tsaregorodtsev \and A. R\"{u}hm \and T. Keller \and L. Botty\'{a}n}

\institute{B. Nagy \and L. F. Kiss \and Sz. Sajti \and D. G. Merkel \and  F. Tanczik\'{o} \and L. Botty\'{a}n \at
              Wigner Research Centre for Physics,\\
              Hungarian Academy of Sciences, Budapest, Hungary\\
              \email{nagy.bela@wigner.mta.hu}
           \and
              Yu. N. Khaydukov \and T. Keller \at
              Max-Planck Institute for Solid State Research, Stuttgart, Germany
           \and
              Yu. N. Khaydukov \and T. Keller \and A. R\"{u}hm \at
              Forschungs-Neutronenquelle Heinz Maier-Leibnitz,\\
              Technische Universit\"{a}t M\"{u}nchen, Munich, Germany
           \and
              Yu. N. Khaydukov \at
              Skobeltsyn Institute of Nuclear Physics,\\ Moscow State University, Moscow, Russia
           \and
              A. S. Vasenko \at
              Institut Laue-Langevin, Grenoble, France
           \and
               R. O. Tsaregorodtsev \at
               Faculty of Physics, Moscow State University, Moscow, Russia
           \and
               A. R\"{u}hm \at
               Max-Planck Institute for Intelligent Systems, Stuttgart, Germany
}

\date{Received: date / Accepted: date}

\maketitle

\begin{abstract}
Thickness ($d_F$) and concentration ($x$) dependence of the Curie temperature of Ni$_x$Cu$_{100-x}$($d_F$)
ferromagnetic (F) alloy layers ($x = 55, 65, d_F$ = [3nm $\div$ 12nm]) being in contact with a vanadium
layer was studied. The Curie temperature of the ferromagnetic layers depends on the thickness when it is
comparable with the interface layer between the F and the vanadium layers, which is attributed to the
proximity coupling of the interface region with the rest of the F layer. The present study provides valuable
information for fabrication of samples with controlled exchange coupling strength for studies of
superconductor/ ferromagnet (S/F) proximity effects.
\end{abstract}


\section{Introduction}

Proximity effects occurring at the interface of superconducting and ferromagnetic phases have attracted
considerable attention in recent years \cite{1}. Extensive theoretical work has been exerted to understand
the nature of these mixed states. For weak ferromagnets, with Curie temperatures below $T_C \sim 100$ K a
transition of the homogeneous F layer into a non-homogeneous domain-like (``cryptoferromagnetic'') phase
is predicted \cite{2}, \cite{3}, \cite{4} with domain sizes less than the coherence length of the superconductor.
Conventional 3d ferromagnets, like Fe, Co and Ni, exhibit Curie temperatures above 300 K. In order to study S/F
proximity effects with a weak ferromagnet, a control of the Curie temperature of the F layer is required. In this
work we use the dilution of a conventional ferromagnetic material with non-magnetic atoms, namely nickel with copper.
Ni and Cu are fully miscible in the bulk \cite{5} with almost linear dependence of the Curie temperature
(and, hence, the exchange coupling strength) on the Ni concentration \cite{6}.

In addition to the concentration dependence of the Curie temperature, the dependence on the thickness of thin F layer
was reported in several works, e.g. $T_{C}(d_{\mathrm{Gd}})$ for Nb/Gd structures \cite{7}. A linear dependence on
$d_\mathrm{Gd}$ was observed for small $d_\mathrm{Gd}$-s, then it is saturated for larger Gd thicknesses, a phe\-no\-me\-non,
which was ascribed to the finite-size effects \cite{8}. In \cite{9} the $T_C(d_F)$ was studied in thin Ni
and Ni$_{60}$Cu$_{40}$ layers being in contact with a Nb layer, and only a linear thickness dependence was observed.

Here we study the magnetic properties of Ni$_x$Cu$_{100-x}$ films ($x=55$ and 65 and $d_F$ =[3nm $\div$ 12nm])
being in contact with a 80nm thick vanadium layer. These systems were prepared to reveal proximity effects between
the superconducting vanadium and the ferromagnetic Ni$_x$Cu$_{100-x}$ and study their dependence on the exchange
coupling strength of the F layer by means of polarized neutron reflectometry. To enhance the contribution of the
V/Ni$_x$Cu$_{100-x}$ interface to the scattered neutron intensity the V layer was covered with a copper layer to
form a neutron waveguide structure \cite{10}. We found the Curie temperature dependence on the F layer thickness,
similarly to \cite{7}: a linear dependence at small enough thicknesses and a saturation at larger thicknesses. This
behavior is ascribed to a proximity coupling of the V/Ni$_x$Cu$_{100-x}$ interface transition region (``dead layer'')
with the rest of the F layer. Although the transition layer is non-magnetic, proximity to the main (non-mixed) part
of the ferromagnetic layer increasingly suppresses the Curie temperature with decreasing the F layer thickness.


\section{Sample preparation and experiment}

Samples with a nominal composition of Cu(31 nm)/ V(80 nm)/ Ni$_x$Cu$_{100-x}$($d_F$)/ MgO
were prepared using Molecular Beam Epitaxy in the Wigner Research Centre for Physics
(Budapest, Hungary). The MgO substrate surface was clea\-ned by rinsing it in the isopropanol
in an ultrasonic cleaner and by heating it to 600 C in ultrahigh vacuum for 30 minutes.
The metallic layers were deposited at room temperature in a base pressure of 5$\times 10^{-11}$ mbar.
All samples were rotated during deposition to increase lateral homogeneity of the films.
Mixing of Ni and Cu was achieved by co-eva\-po\-ra\-tion with deposition rates
$\nu_\mathrm{Ni} = 0.11 {\mathrm\AA}$/s and $\nu_\mathrm{Cu} = 0.1 {\mathrm\AA}$/s for $x = 55$
and $\nu_\mathrm{Ni} = 0.17 {\mathrm\AA}$/s, $\nu_\mathrm{Cu} = 0.1 {\mathrm\AA}$/s for $x= 65$.

\begin{figure*}[t]\begin{center}
\includegraphics[scale=0.3]{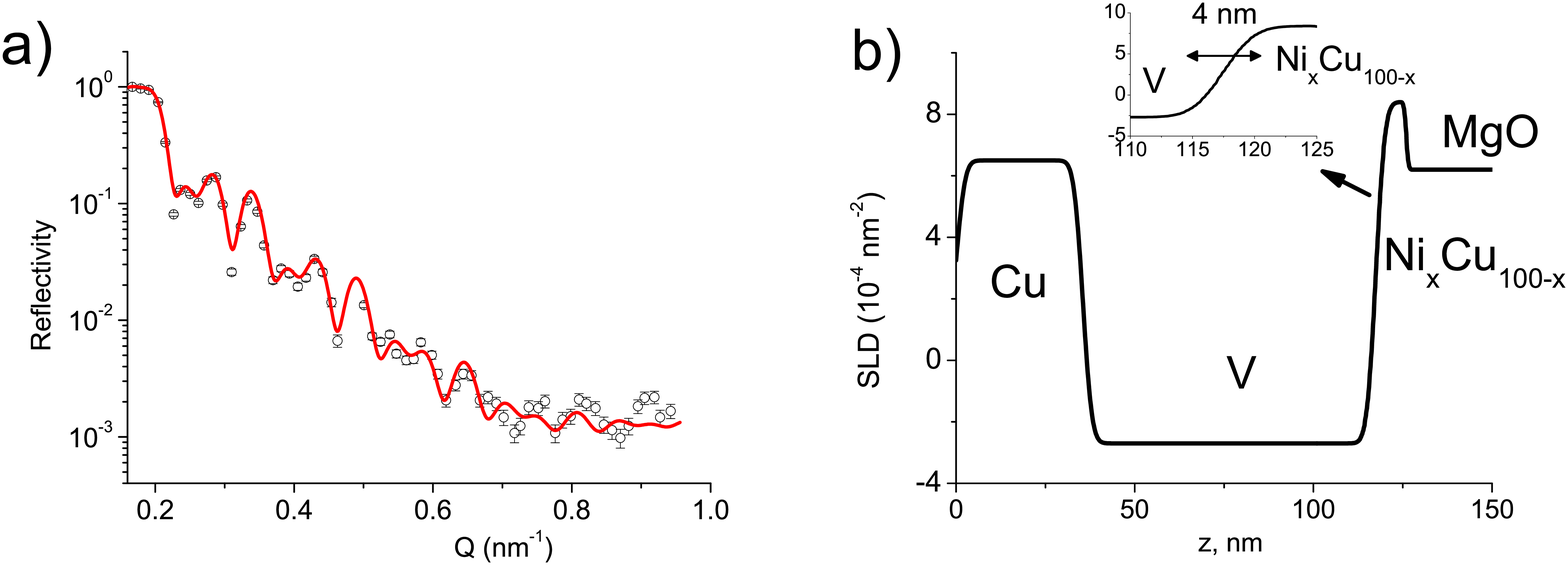}
\caption{Experimental (dots) and model (solid line) reflectivity curves measured on sample
Cu(31nm)/ V(80nm)/ Ni$_{65}$Cu$_{35}$(8.5nm)/ MgO, (a). Extracted depth-profile of nuclear
scattering length density of entire structure, (b). Inset: SLD profile around the V/Ni$_x$Cu$_{100-x}$
interface can be characterized by the presence of a transition layer with a thickness of the order of 4 nm.}
\label{Fig1} \vspace{-4mm}
\end{center}
\end{figure*}

The layer structure of the samples was characterized by neutron reflectometry at angle-dispersive
neutron reflectometers GINA (Budapest Neutron Center, Hungary) and NREX (research reactor FRM-II,
Garching, Germany). The measurements at the GINA reflectometer were conducted in $H = 500$ Oe external
field at room temperature in a momentum transfer range of $Q = [0.08 \div 0.95]$ nm$^{-1}$. At
room temperature the spin-up and spin-down reflectivities were identical and the sum of reflectograms
in the two polarizations was fitted. A typical neutron reflectivity curve measured on the sample
Cu(31 nm)/ V(80 nm)/ Ni$_{65}$Cu$_{35}$(8.5 nm)/ MgO is shown in Fig.~\ref{Fig1}a.

The reflectivity curve is characterized by Kiessig oscillations due to the interference on different
interfaces. Fit of the data by the program FitSuite \cite{11} allowed to restore the nuclear scattering
length density (SLD) profiles (Fig.~\ref{Fig1}b). From these profiles the actual thicknesses of F layers
were derived (Table 1). In addition, the analysis of neutron
reflectometry data suggests the existence of non-abrupt interface between vanadium and F layers, with
a typical thickness between 2-5 nm (see inset in the Fig.~\ref{Fig1}b).

\begin{figure*}[t]\begin{center}
\includegraphics[scale=0.31]{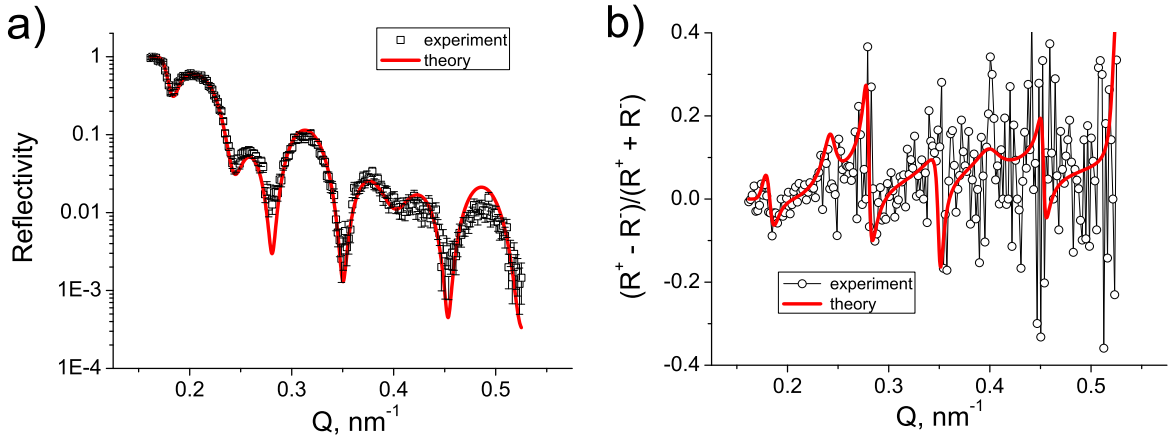}
\vspace{4mm}
\caption{$R^+$ reflectivity from the sample Cu(31nm)/ V(80nm)/ Ni$_{55}$Cu$_{45}$(6nm)/ MgO measured at $T = 10$K
in $H = 200$ Oe external field, (a). The solid line corresponds to the model explained in the text. Experimental
(open circles) and model (full line) spin asymmetries $(R^+-R^-)/(R^++R^-)$ corresponding to the nuclear
SLD profile and magnetization $4 M = 1.4$ kGs in the Ni$_{55}$Cu$_{45}$(6nm) layer.}
\label{Fig2} \vspace{-4mm}
\end{center}
\end{figure*}

A set of low-temperature PNR measurements have been performed on the Cu(31 nm)/ V(80 nm)/ Ni$_{55}$Cu$_{45}$(6 nm)/ MgO
sample at the NREX neutron reflectometer. Magnetic field $H = 200$ Oe was applied parallel to the surface.
Spin-up and spin-down reflectivities $R^+(Q)$ and $R^-(Q)$ were measured in a momentum transfer range of
$Q = [0.17 \div 0.5]$ nm$^{-1}$ (Fig.~\ref{Fig2}a). Those reflectivities could also be well described using
the SLD profiles obtained from the fitting of the room temperature data.


In order to quantify the magnetic moment of the F layer, the spin asymmetry
[the normalized difference of the spin-up and spin-down reflectivities
($R^+(Q)$ - $R^-(Q)$)/($R^+(Q)$ + $R^-(Q)$), Fig.~\ref{Fig2}b] was modeled. The full line in Fig.~\ref{Fig2}b
corresponds to the above-described SLD profile and a $4 M = 1.4 \pm 0.2$ kGs magnetization
in the Ni$_{55}$Cu$_{45}$ layer, which is slightly higher than the 1.1 kGs measured by Rusanov \textit{et al.}
\cite{12} for a similar Ni concentration.

\begin{figure*}[t]\begin{center}
\includegraphics[scale=0.52]{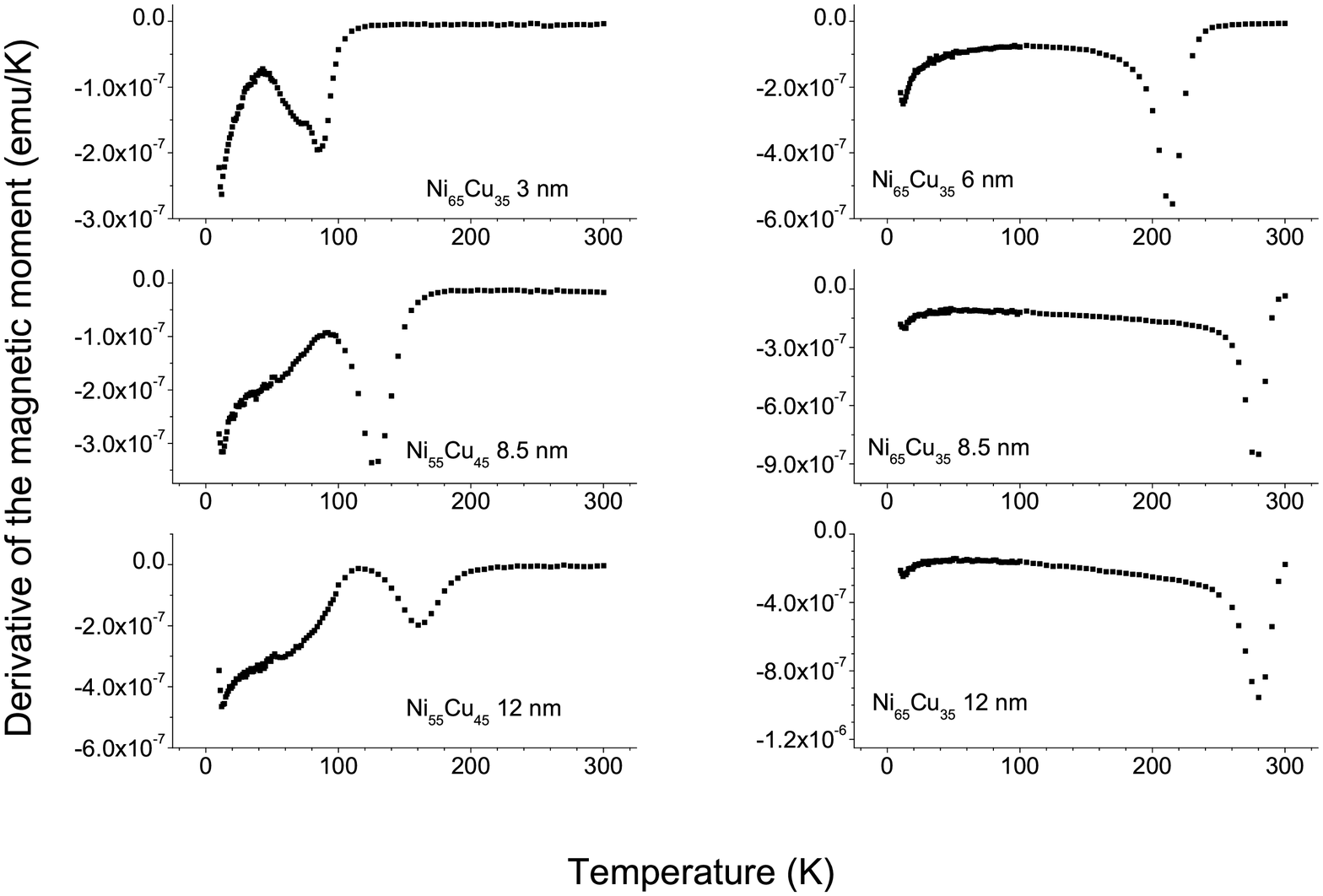}
\vspace{4mm}
\caption{Temperature dependence of the derivative of the magnetic moment of the samples
measured by SQUID. Curie temperatures identified as the temperature of the minimum of
the derivative magnetic moment are listed in Table 1.} \label{Fig3} \vspace{-4mm}
\end{center}
\end{figure*}

The Curie temperature of the samples was determined from data collected by the
(Quantum Design) SQUID magnetometer at the Wigner Research Center, Budapest.
Following a cooling down in zero external magnetic field, the temperature dependence
of the magnetic moment of the samples was measured in 10 Oe magnetic field applied
parallel with the sample surface (Fig.~\ref{Fig3}). The Curie temperature was defined
as the temperature of the minimum in the derivative of the magnetic moment curve (Fig.~\ref{Fig3} and Table 1).

\vspace{4mm}
\begin{center}
\begin{tabular}{|c|c|c|}
\hline
\multicolumn{3}{|c|}{Table 1. Fitted F layer thicknesses}
\\
\multicolumn{3}{|c|}{and measured Curie temperatures.}
\\ \hline
\multirow{2}{*}
{Sample} & Fitted F-layer & Curie \\
& thickness [nm] & temperature [K]
\\ \hline
Ni$_{65}$Cu$_{35}$ 3 nm & 4.5 & 86
\\ \hline
Ni$_{65}$Cu$_{35}$ 6 nm & 7.0 & 215
\\ \hline
Ni$_{65}$Cu$_{35}$ 8.5 nm  &  8.5 & 280
\\ \hline
Ni$_{65}$Cu$_{35}$ 12 nm  &  12.7 & 280
\\ \hline
Ni$_{55}$Cu$_{45}$ 8.5 nm  &  6.0 & 130
\\ \hline
Ni$_{55}$Cu$_{45}$ 12 nm  &  - & 160
\\ \hline
\end{tabular}
\end{center}


\section{Results and discussion}

\begin{figure}[h]\begin{center}
\includegraphics[scale=0.32]{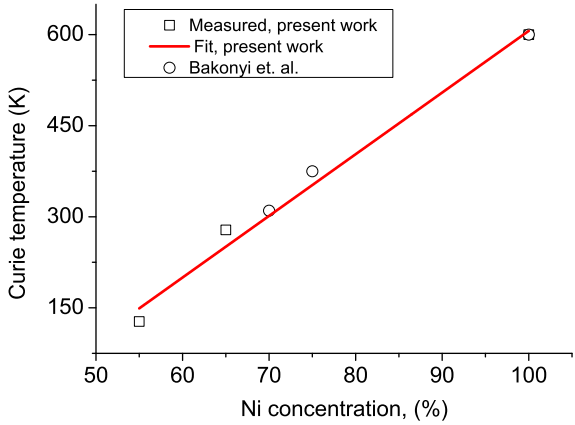}
\vspace{4mm}
\caption{Ni-concentration dependence of the Curie temperature for
Ni$_x$Cu$_{100-x}$ layers with F-layer thickness of $d_F = 8.5$ nm (open squares)
along with literature data from Bakonyi \textit{et al.} \cite{6}. } \label{Fig4}
\vspace{-4mm}
\end{center}
\end{figure}
\begin{figure}[h]\begin{center}
\includegraphics[scale=0.32]{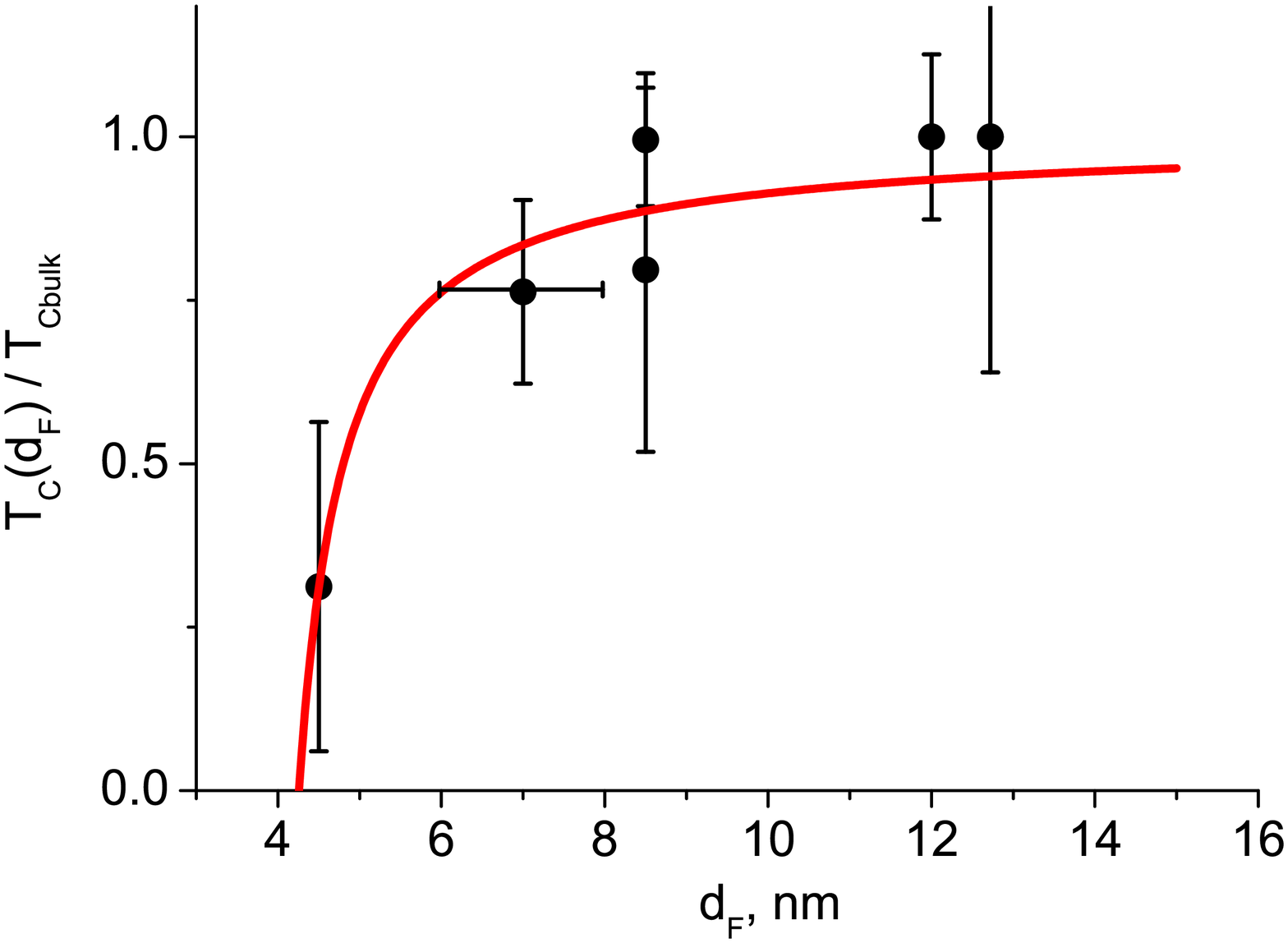}
\vspace{4mm}
\caption{Dependence of the Curie temperature on the thickness of the F layer and simulation
according to expression \eqref{2} with $d_T = 4$ nm and $c = 0.1$.} \label{Fig5}
\vspace{-4mm}
\end{center}
\end{figure}

Dependence of the Curie temperature on the Ni concentration for samples with $d_F = 8.5$ nm is shown
in Fig.~\ref{Fig4} along with data from \cite{6}. This dependence is linear and agrees
with the behavior of bulk Ni$_x$Cu$_{100-x}$ alloys for $x > 55$ \cite{6}. The linear fit
to the present data provides $T_C = 0$ K for a Ni concentration $x = 40 \pm 5 \%$.
This allows us to conclude that samples with $d_F > 8.5$ nm behave like the bulk alloy.
However, we observed a dependence of the Curie temperature on the F layer thickness (Fig.~\ref{Fig5}),
similar to the one reported in \cite{7} for Nb/Gd structures. Namely, the Curie temperature increases
with increasing the F layer thickness for samples with $d_F < 8.5$ nm and then approaches a saturation
corresponding to the bulk value. While in \cite{7} this behavior was attributed to the manifestation
of the finite-size effects, we ascribe the thickness dependence of the Curie temperature to the influence
of the interface of vanadium and Ni$_x$Cu$_{100-x}$ layers. Indeed, neutron reflectometry gives estimation for the thickness
of the transition (T) layer between vanadium and Ni$_x$Cu$_{100-x}$ of the order of $d_T \sim$ 2-5 nm. This transition
layer, due to its vanadium content is mostly non-magnetic, but proximity to the main part of the F
layer (M) may lead to an induced magnetization in it. In other words we model the F layer as a proximity-coupled
T/M bilayer.

Following Bergeret \textit{et al.} \cite{13} we can write the expression for the
effective exchange field $H_\mathrm{eff}$ of the T/M bilayer as,
\begin{equation}\label{1}
H_\mathrm{eff} = \frac{H \nu_M d_M}{\nu_M d_M + \nu_T d_T},
\end{equation}
where $H$ is the exchange field of the bulk Ni$_x$Cu$_{100-x}$, $\nu_{M,T}$ and $d_{M,T}$ are the densities of states
(DOS) and thicknesses of the main part of the F layer and the transition layer, respectively ($d_F \equiv d_M + d_T$).
Eq.~\eqref{1} is valid provided the thicknesses of M and T layers are smaller than the spin diffusion length,
i.e. the spin is conserved in the samples. Postulating proportionality between exchange field and the Curie
temperature we can rewrite \eqref{1} as,
\begin{equation}\label{2}
\frac{T_C(d_F)}{T_{C \mathrm{bulk}}} = \frac{d_M}{d_M + c d_T},
\end{equation}
where $T_{C \mathrm{bulk}}$ is the bulk Curie temperature for the given concentration \cite{6} and
$c = \nu_T / \nu_M$  is the ratio of DOS of the T and M layers.

Eq.~\eqref{2} can qualitatively explain
the thickness dependence of the Curie temperature. Indeed, when the thickness of the transition layer
is comparable with the total thickness of the F layer the dependence $T_C(d_F)$ is almost linear.
When the thickness of the F layer is much larger than the thickness of the transition layer, the
Curie temperature of the layer is close to the bulk value. Fit of the data to the expression \eqref{2}
provides $d_T = 4$ nm, a value comparable with the thickness of the transition layer
estimated from the neutron data, and $c = 0.1$. Similar values of the non-magnetic transition layer (or the ``dead layer''
as it is often called in the literature) for Ni$_x$Cu$_{100-x}$ alloys were reported in other experiments
like measurements of oscillations of the critical current in S/F/S Josephson junctions \cite{14} or
direct measurements of the DOS \cite{15}. Since the transition layer is non-magnetic, it does not play
a role in the ``oscillating'' superconductivity \cite{16}. The presence of a transition layer of
the same order of 1 nm was similarly reported for Nb/PdNi structures \cite{17}. The relatively small
value $c = 0.1$ indicates that the transition layer exhibits a much smaller DOS than the one in the
ferromagnetic layer. Such a large decrease in the DOS indicate a change in the character of the
interface region from a majority $d-$like to a majority $s-$like, i.e. to a non-magnetic one.

Moreover, interface mixing considerably increases the resistance of the V/Ni$_x$Cu$_{100-x}$ interface,
similar to what has been reported for Nb/Ni$_{60}$Cu$_{40}$ structures in \cite{9}. Indeed, both V and Nb
represent strong scattering centers in both Cu and Ni, resulting in specific residual resistivities 5 to 6
times higher than does Cu in Ni or Ni in Cu \cite{18}. The interface resistance and its relation to the
interface transparency in S/weak-F hybrid structures was elaborated in \cite{19}, \cite{20}.

\section{Conclusions}

In conclusion, thickness and concentration dependencies of the Curie temperature of
Ni$_x$Cu$_{100-x}$ ferromagnetic layer being in contact with a vanadium layer was studied.
The Curie temperature was found to increase with the thickness of the F layer when the thickness
is comparable to that of the mixed V/F interface region and it is saturated at larger thicknesses
to the bulk value. The dependence is explained by the proximity coupling of the mixed transition
layer with the rest of the F layer. The present study provides valuable information for fabrication
of samples with controlled exchange coupling strength for studies of the superconductor/ ferromagnet
proximity effects.


\begin{acknowledgements}
The authors are grateful to Mr. G. Kert\'{e}sz for his help in the sample preparation and
to Drs. J. Major, I. Bakonyi and A. A. Golubov for the fruitful discussions. Financial
support by the National Office of Innovation of Hungary under contract NAP-VENEUS, and
by the European Commission under the 7$^{th}$ Framework Programme through the `Research
Infrastructures' action, NMI3 are gratefully acknowledged.
\end{acknowledgements}

\end{document}